\documentstyle[aps,twocolumn,epsf]{revtex}

\begin{document}

\draft

\title{
Cooling of a single atom in an optical trap inside a resonator}
\author{S.J. van Enk$^1$,
J. McKeever$^1$, H.J. Kimble$^1$ and J. Ye$^2$}
\address{$^1$Norman Bridge Laboratory of Physics, California
Institute of Technology 12-33, Pasadena, CA 91125\\ $^2$JILA, Campus Box 440,
 University of Colorado and
National Institute of Standards and Technology Boulder, CO 80309-0440}

\date{\today}
\maketitle

\begin{abstract}
We present detailed discussions of cooling and trapping mechanisms for an
atom in an optical trap inside an optical cavity, as relevant to recent
experiments. The interference pattern of cavity QED and trapping fields in
space makes the trapping wells in principle distinguishable from one another. This adds
considerable flexibility  to creating effective trapping and cooling
conditions and to detection possibilities. Friction and diffusion
coefficients are calculated in and beyond the low excitation limit and full
3-D simulations of the quasiclassical motion of a Cs atom are performed.
\end{abstract}
\pacs{32.80.Pj, 42.50.Vk, 42.50.Lc}
\section{Introduction}
A recent experiment \cite{hood} succeeded in trapping a single atom with
single photons inside an optical cavity and in monitoring the atomic motion
with the resolution approaching the standard quantum limit for position
measurements. Yet a second experiment \cite{pinkse} has likewise reported
single-atom trapping at the few-photon level, although in this case the
trapping potential and diffusion are in fact well approximated by a
free-space semiclassical theory \cite{doh}.

 One future objective for such experiments is to use atoms trapped in cavities for
quantum communication purposes, with atoms serving as quantum memories and
photons as the transporters of quantum information \cite{qc,qc2}. While the
single-photon trapping experiments provide a new paradigm for quantum
measurement and control, they are nevertheless not entirely suitable for the
purpose of distributed quantum networks where qubits will be communicated
among quantum nodes. The reason is the short trapping life time of the atoms
as well as limited operation flexibility. A better strategy might be to use
the cavity QED field for quantum state entanglement and distribution while an
additional (external) trapping mechanism provides the necessary confinement
of the atomic center-of-mass motion.  For instance, in another recent
experiment from the Caltech group \cite{ye}, mean trapping times of $\sim
28$ms (as compared to mean trapping times of $< 1$msec in the experiments
\cite{hood,pinkse}) were achieved by employing a far-off resonant trapping
(FORT) beam along the cavity axis. In that experiment the trapping lifetime
was limited due to intensity fluctuations of the intracavity FORT beam
\cite{gardiner}. Here we consider the situation of current improved
experiments \cite{exp} in which a single atom is held inside an optical
cavity in a stable FORT beam of minimum intensity fluctuations.

Several mechanisms for cooling inside optical resonators have been discussed
before \cite{mossberg,horak,chu}. Here we discuss in detail how the
combination of an external trapping potential and the cavity QED field adds
flexibility in predetermining where and to what degree atoms will be trapped
and cooled. Moreover, our calculations go beyond the weak driving limit
discussed in \cite{horak}. That is, we allow the ``probe'' field driving the
cavity to be so strong as to appreciably modify the dynamical behavior of,
rather than merely probe, the atom-cavity system.

This paper is organized as follows. In Section II we describe the physical
situation of an atom trapped in an optical potential and strongly interacting
with a cavity QED field. We give the evolution equations for both internal
and external atomic degrees of freedom and for the quantized cavity mode.
Section III contains an exposition on how we calculated friction and
diffusion coefficients from the forces acting on the atom. Section IV
contains the main results of this paper: we discuss simple pictures for
cooling mechanisms, based on the dressed state structure of the atom-cavity
system, and give numerical results for the typical cooling and diffusion
rates, and hence ``temperatures'' for single atoms under various trapping
conditions. We also study the saturation behavior under strong driving
conditions and perform simulations of the full 3-D motion of atoms trapped in
particular wells that show how the probe field transmission is correlated
with the atomic motion and how trapping times can be prolonged by strong
cooling. Section \ref{caseB} concludes with a brief discussion of a slightly
different trapping scheme. The summary highlights the main results.
\section{Description of problem}
 We consider a single two-level atom coupled to a single quantized cavity
mode and coupled to a (classical) far-off resonant trapping (FORT) beam. In
most of the paper we assume that the FORT shifts the atomic excited state
$|e\rangle$ {\em up} and the ground state $|g\rangle$ {\em down} by an amount
$S_F(\vec{r})$ (i.e., the energy of the ground state is $E_g-S_F$, that of
the excited state $E_e+S_F$), as this is the situation pursued in previous
and current experiments \cite{ye,exp}. In Section \ref{caseB}, however, we
will also study the different situation where both ground and excited states
are shifted {\em down} by $S_F$ (see e.g. \cite{icols}). The FORT beam
coincides with one of the longitudinal modes of the cavity and its wavelength
$\lambda_F$ is longer than that of the main cavity mode of interest for
cavity QED, $\lambda_0$. In fact, in the experiments \cite{ye,exp} the cavity
length $L$ is $104\lambda_0/2=102\lambda_F/2$.

\begin{figure}[tbp]
 \leavevmode    \epsfxsize=8cm
\epsfbox{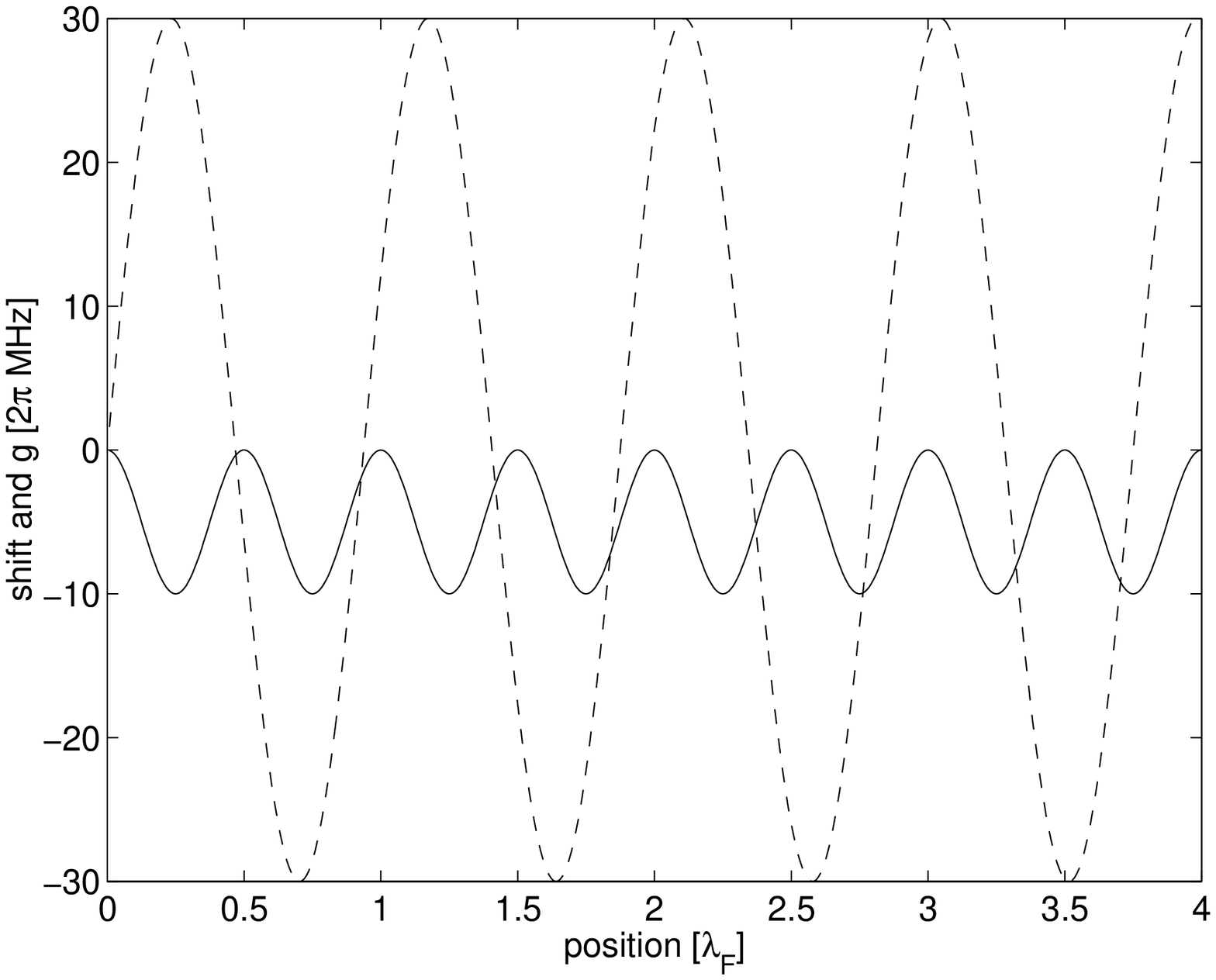} \caption{The FORT-induced  shift of the ground state on
axis ($\rho=0$) in the case where $S_0/(2\pi)=10$MHz and the cavity-QED
coupling rate $g$ (dashed curve), with $g_0=3S_0$, as functions of position
along the cavity axis measured in units of the FORT wavelength $\lambda_F$,
with $z=0$ at the left cavity boundary. The cavity length is $L=15\lambda_F$.
}\label{shiftg}
\end{figure}

\begin{figure} [tbp]
 \leavevmode    \epsfxsize=8cm
\epsfbox{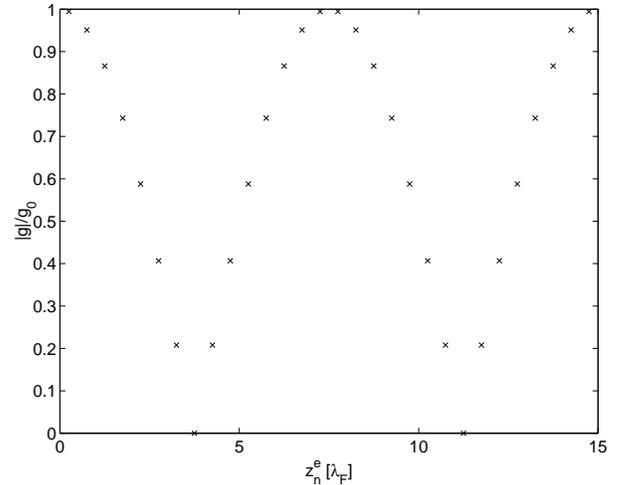} \caption{The values of $|g|/g_0$ at the locations of the
anti-nodes of the FORT, i.e. at the points $z_n=(n/2-1/4)\lambda_F$ for
$n=1\ldots 30$. There are 8 quantitatively different wells.}\label{beatgF}
\end{figure}

 The position-dependent AC-Stark shift due to the FORT
field is  of the form
\begin{equation}
S_F(\vec{r})=S_0\sin^2(k_F z)\exp(-2\rho^2/w_0^2),
\end{equation}
with $S_0>0$ the maximum shift, $k_F=2\pi/\lambda_F$ the wave vector of the
FORT field, $w_0$ the size of the Gaussian mode of the cavity, while $z$ and
$\rho$ give the coordinate along, and the distance perpendicular to, the
cavity axis, respectively. The quantized cavity mode is assumed to have the
same transverse dimensions $w_0$\footnote{It is in fact the Rayleigh ranges
of the beams that are identical, so that $w_0^{{\rm FORT}}/w_0^{{\rm cav}}=
\sqrt{\lambda_0/\lambda_F}\approx 0.99$} so that the atom-cavity coupling is
determined by
\begin{equation}
g(\vec{r})=g_0\sin(kz)\exp(-\rho^2/w_0^2),
\end{equation}
with $g_0$ the maximum coupling rate and $k=2\pi/\lambda_0$ the wave vector
of the cavity mode. Under conditions where the cavity is not driven too
strongly, the atom will be trapped around the anti-nodes of the red-detuned
FORT field. Thanks to the fact that $\lambda_0\neq\lambda_F$, the atom will
experience a different coupling strength to the cavity mode in each different
well. Figure~\ref{shiftg} shows the axial pattern arising from the FORT and
cavity fields.
 For
illustrative purposes we choose here (and in the rest of this paper) a cavity
of length $L=16\lambda_0=15\lambda_F$. This does not influence the basic
physics involved: in particular we note that the precise value of $\lambda_F$
is largely irrelevant on the time scales considered here, as the FORT field
is detuned far from atomic resonance. The choice of
$L=16\lambda_0=15\lambda_F$ just means that only 8 wells out of 30 are
qualitatively and quantitatively different.
This is illustrated in Fig.~\ref{beatgF} where we plot the value of the
cavity QED coupling $g$ at the anti-nodes  of the FORT (i.e., the bottom of
the trapping potential). In particular, there are 2 anti-nodes in which
$g=0$, and 4 in which $|g|$ attains it maximum.

 The cavity is driven by an external
classical field ${\cal E}(t)={\cal E}_0\exp(i\omega_p t)$, at a frequency
$\omega_p$, which is used to probe the atom-cavity system and which may cool
the atom at the same time. In the following, the strength of the driving
field is indicated by the number of cavity photons $N_e$ that would be
present if there were no atom in the cavity, rather than by ${\cal E}_0$.
This closely follows the experimental procedure for determining the driving
strength. The relation between the two is
\begin{equation}
N_e=\frac{{\cal E}_0^2}{\kappa^2+\Delta_c^2},
\end{equation}
with  $\Delta_c=\omega_c-\omega_p$ the detuning of the probe from the cavity
frequency $\omega_c=kc$.
 The Hamiltonian for the internal atomic degrees of freedom and the quantized cavity mode
 is, in a frame rotating at the probe frequency $\omega_p$, given by
\begin{eqnarray}
H&=&\hbar \Delta_c a^+a+\hbar\Delta_a\sigma^+\sigma^- +2\hbar
S_F(\vec{r})(\sigma^+\sigma^--1/2)\nonumber\\
 &&+\hbar {\cal
E}_0(a^++a)+\hbar g(\vec{r})(a^+\sigma^-+\sigma^+a).
\end{eqnarray}
Here $\Delta_a=\omega_a-\omega_p$ is the detuning of the atomic resonance
from the probe frequency. In all numerical examples given below the cavity
frequency is chosen to coincide with the atomic frequency, so that
$\Delta_c=\Delta_a$. The quantity $\Delta_p\equiv -\Delta_a$ is then referred
to as the probe detuning. Note here that without a FORT the optimum cavity and atom
detunings are not equal \cite{mossberg,horak,chu}. In our case, however,
the FORT effectively changes the atomic frequency in a position-dependent way
and thus the precise value of the atomic detuning relative to the cavity detuning is largely irrelevant. Indeed optimum cooling conditions will exist in certain wells but not in others,
which is one feature that allows one to distinguish various wells.
 
Coupling the atom and the cavity to the remaining modes of the
electro-magnetic field leads by a standard procedure to the master equation
for the density operator of the coupled atom-cavity system,
\begin{eqnarray}\label{rho}
\frac{{\rm d} \rho}{{\rm d} t}=-i[H,\rho]/\hbar -\kappa \{a^+a,\rho\}+2\kappa
a\rho a^+ +\nonumber\\ -\frac{\Gamma}{2}\{\sigma^+\sigma^-,\rho\}+
\frac{3\Gamma}{8\pi} \int {\rm d}^2\hat{k}
\sum_{\hat{\epsilon}}(\hat{d}\cdot\hat{\epsilon})^2
\exp(-i\vec{k}\cdot\vec{r})\times\nonumber\\ \times\sigma^-\rho\sigma^+
\exp(i\vec{k}\cdot\vec{r}),
\end{eqnarray}
with $\Gamma$ the spontaneous decay rate and $\kappa$ the cavity decay rate.
We are mainly interested in the strong-coupling regime, where $g_0\gg
\Gamma,\kappa$.

We treat the external (center-of-mass) degrees of freedom of the atom
classically, an approximation justified at the end of Section \ref{cool}. For
a discussion of various interesting effects arising from the quantized
external motion of an atom in a cavity QED field, we refer the reader to
\cite{quant}.

 In the quasiclassical approximation (i.e., where we retain
the full quantum character of the internal degrees of freedom and of the
cavity mode; see \cite{doh} for a full discussion of this approximation), the
integral in (\ref{rho}) can be evaluated to give the simpler result
\begin{eqnarray}\label{rhos}
\frac{{\rm d} \rho}{{\rm d} t}&=&-i[H,\rho]/\hbar -\kappa
\{a^+a,\rho\}+2\kappa a\rho a^+ \nonumber\\
&&-\frac{\Gamma}{2}\{\sigma^+\sigma^-,\rho\}+\Gamma \sigma^-\rho\sigma^+.
\end{eqnarray}
 The force acting on the atom consists of two parts, one due to
spontaneous emission, whose mean vanishes on average, and the other part is
represented by the operator
\begin{eqnarray}
\label{force} \vec{F}&\equiv&-\vec{\nabla}H\nonumber\\ &=&-2\hbar\vec{\nabla}
S_F (\sigma^+\sigma^--1/2) -\hbar\vec{\nabla} g (a^+\sigma^-+\sigma^+a),
\end{eqnarray}
which has contributions arising from the FORT potential and from the
interaction with the cavity mode. It was only the latter part that was
considered in \cite{horak} and that leads to 1-D cooling to temperatures of
the order $k_BT\sim {\rm min}(\hbar\kappa,\hbar\Gamma/2)$. See also
Refs.~\cite{doherty} for similar calculations on single atoms moving in
cavity QED field, and Refs.~\cite{taieb,marksteiner} for calculations of
diffusion of atoms in optical traps in free space.

It can be shown \cite{coh} starting from a fully quantized description, that
the semiclassical motion of the atom is described by a Fokker-Planck equation
for the Wigner distribution function containing (position-dependent) friction and diffusion
coefficients. Equivalently, we may use stochastic equations for the classical
atomic position and velocity variables $\vec{r}$ and $\vec{v}$ of the form
\begin{eqnarray}\label{lang}
{\rm d}\vec{r}&=&\vec{v}{\rm d}t,\nonumber\\ 
{\rm d}\vec{v} &=& \frac{\langle\vec{F}\rangle}{m} {\rm d}t
-\beta \vec{v} {\rm d}t+ B{\rm d}\vec{W},
\end{eqnarray}
where $\langle.\rangle$ denotes an expectation value, $\beta$ is the friction
tensor (with dimensions of a rate), $m$ the mass of the atom, $B$ is a tensor
such that $D=BB^T/2$ is the velocity diffusion tensor (with dimension
m$^2$/s$^3$), and ${\rm d}\vec{W}$ is a 3-dimensional Wiener process which
satisfies ${\rm d}W_i{\rm d}W_j=\delta_{ij}{\rm d}t$ \cite{gardinerbook}.
Starting with the expression (\ref{force}) for the force operator, we can
calculate $\beta$ and $D$ by the procedure outlined in the next Section.
\section{Friction and diffusion}
Refs.~\cite{horak} employ Heisenberg equations of motion for various field
and atomic operators to find friction and diffusion coefficients. These
equations are not closed and, consequently, an approximation has to be made
in order to find solutions. The natural assumption is to consider the weak
driving limit (i.e., ${\cal E}_0\ll \kappa$) and truncate the available
Hilbert space to that part containing no more than a single cavity photon.
This allows one to close the Heisenberg equations \cite{horak}. Here we
employ a different method (using the density matrix equations) to calculate
friction and diffusion coefficients that does not require us to stay within
the weak driving limit, but in addition  we used Ref.~\cite{horak}'s
procedure here to obtain results in the weak driving limit for verification
purposes. In any case, it is still true that the most interesting regime is
where only one or few photons are involved. Note that
 given the strong coupling
between atom and cavity field, even a single photon is sufficient to lead to
regimes far beyond the weak driving limit. In our examples we truncated the
Hilbert space to photon numbers of around 4 or smaller. We refer to
\cite{tan} for an exposition on how to represent operators in truncated
Hilbert spaces of precisely this form in a numerically convenient manner.

The master equation (\ref{rhos}) is written as
\begin{equation}\label{L}
\frac{{\rm d} \rho}{{\rm d} t}={\cal L} \rho.
\end{equation}
Numerically, the Liouvillian superoperator ${\cal L}$ is converted into a
pre-multiplication operator by methods explained in \cite{tan}. In order to
find friction and diffusion coefficients we apply a simple procedure, which
yields these coefficients at zero velocity: this is sufficient for our
purposes as the atom we are interested in, Cs, is  relatively heavy. More
precisely, the relevant dimensionless parameters determining the velocity
dependence of friction and diffusion coefficients are $kv/\Gamma$ and
$kv/\kappa$ (see for instance \cite{moelmer}), and both are very small in all
our simulations. In particular, $\Gamma/k\sim 4.3$ m/s and $\kappa/k\sim 3.4$
m/s, while velocities in the trapping regime we are interested in (where
atoms are localized in wells at low temperatures for times
$\gg\kappa^{-1},\Gamma^{-1}$) are around the Doppler limit velocity
\begin{equation}
v_D=\sqrt{\frac{\hbar\Gamma/2}{m}}\approx 8.8{\rm cm/s}.
\end{equation}
 Also note that the
standard procedure of continued fractions to calculate the full velocity
dependence is not directly applicable to the present case, as the potential
through which the atom is moving is not periodic ($\lambda_F\neq\lambda_0$).

For an atom moving at velocity $\vec{v}$ we write
\begin{equation}
\frac{{\rm d}}{{\rm d} t}=
 \frac{\partial }{\partial t}+\vec{v}\cdot\nabla,
\end{equation}
and expand (\ref{L}) in powers in $\vec{v}$ and solve for the steady state.
The zeroth-order solution is then the steady state $\rho_0$ at zero velocity:
\begin{equation}\label{0}
{\cal L} \rho_0=0,
\end{equation}
while the first-order term $\rho_1$ is determined by
\begin{equation}\label{1}
{\cal L}\rho_1=\vec{v}\cdot\nabla \rho_0.
\end{equation}
The zeroth-order force is the steady-state force for an atom at rest, and is
given by
\begin{equation}
\vec{F}_0=-{\rm Tr} (\rho_0 \vec{\nabla} H).
\end{equation}
Similarly, the friction coefficients follow from the first-order term in the
force
\begin{equation}\label{F1}
\vec{F}_1=-{\rm Tr} (\rho_1 \vec{\nabla} H),
\end{equation}
by identifying
\begin{equation}
\vec{F}_1\equiv -\beta m\vec{v},
\end{equation}
where $\beta$ is  a 3-by-3 tensor. In our case (\cite{ye}), the gradients
along the cavity axis are larger in magnitude than those in the transverse
directions by roughly a factor $k w_0\approx 150$ (and around the cavity axis
where the atoms spend most of their time the radial gradients are even
smaller, of course). Since the friction coefficient scales with the product
of two gradients (cf.~Eqs (\ref{1}) and (\ref{F1})), the largest element of
the tensor $\beta$ is the $zz$ component. Next largest in magnitude are the
off-diagonal components such as $\beta_{xz}$ and $\beta_{zx}$. Their effects,
however, can be safely neglected in our case: firstly, the force in the $z$
direction proportional to $-\beta_{zx}v_x$ is smaller than the friction force
$-\beta_{zz}v_z$ by roughly a factor $kw_0$. Secondly, the force in the $x$
direction $-\beta_{xz}v_z$ is not a friction force (as it is not proportional
to $v_x$), and its contribution is averaged out because the oscillations in
$v_z$ are faster than those in the $x$ direction by another factor $kw_0$.
Finally, the purely radial friction rates such as $\beta_{xx}$ are too small
($\ll 1$s$^{-1}$ on average) to have any influence on the time scales
considered here. Thus we take only $\beta_{zz}$ into account.

 The diffusion coefficient,
again at zero velocity, is calculated as follows. The standard method is to
use the quantum regression theorem, and a particularly useful (for numerical
purposes) interpretation of that theorem is given in \cite{tan}. The momentum
diffusion tensor $D_p$ is given by
\begin{equation}\label{D}
D_p=\lim_{t\rightarrow\infty}{\rm Re} \int _0^\infty {\rm d} \tau \langle
\vec{F}(t) \vec{F}(t-\tau)\rangle -\langle \vec{F}(t)\rangle \langle
\vec{F}(t-\tau)\rangle,
\end{equation}
and its relation to the velocity diffusion tensor is $D=D_p/m^2$.
 Before
eliminating any degrees of freedom, the total system in fully quantized form
is described by a time-independent Hamiltonian, which we denote by $H_{{\rm
tot}}$. In that case the time evolution of all operators is determined by
$\exp(-iH_{{\rm tot}}t)$, and two-time averages of the form $\langle A(t)
B(t-\tau) \rangle$ as appearing in (\ref{D}) can be written as
\begin{equation}\label{AB}
\langle A(t) B(t-\tau) \rangle= {\rm Tr} \left[A \exp(-iH_{{\rm tot}}\tau) B
\rho_{{\rm tot}}(t) \exp(iH_{{\rm tot}}\tau)\right],
\end{equation}
with $\rho_{{\rm tot}}$ the density matrix of the total system. This
expression formally contains the evolution of a density matrix over a time
interval $\tau$ starting from an initial density matrix $\rho_{{\rm
init}}\equiv B \rho_{{\rm tot}}(t)$. The quantum regression theorem now
states that (\ref{AB}) is still valid for the reduced density matrix that
evolves under the Liouvillian ${\cal L}$. That is, instead of (\ref{AB}) we
may use
\begin{equation}\label{ABr}
\langle A(t) B(t-\tau) \rangle= {\rm Tr} \left[A \exp({\cal L}\tau) B
\rho(t)\right].
\end{equation}
In our case, ${\cal L}$ is a time-independent operator and hence the
right-hand side of (\ref{ABr}) can be evaluated by expanding $\exp({\cal
L}\tau)$ in an exponential time series, as in the methods developed in
\cite{tan}. This then is the method we use here to evaluate the friction and
diffusion tensors, and the results have been checked in the low-intensity
limit by applying the different methods from \cite{horak} to the same
problem.

Diffusion due to spontaneous emission is not obtained this way (as the bath
of vacuum modes has been eliminated already), but can be obtained by standard
methods and gives an independent additional three components
$(D_p)_{ii}^{SE}=N_{i}\hbar^2k^2\Gamma/2\langle\sigma^+\sigma^-\rangle_0$ for
$i=x,y,z$, with $\langle.\rangle_0$ denoting a steady-state value and with
the dimensionless factor $N_i$ depending on polarization. When the two-level
system is formed by  two Zeeman levels that are connected by circularly
polarized light propagating in the $z$ direction, we have $N_z=2/5$, and
$N_x=N_y=3/10$.

\begin{figure}[tbp]
 \leavevmode    \epsfxsize=8cm
\epsfbox{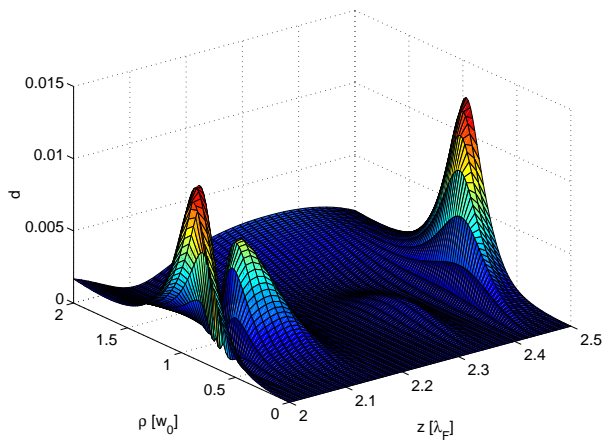} \caption{For parameters to be used later,
$S_0/(2\pi)=50$MHz, and $\Delta_p/(2\pi)=-10$MHz, $N_e=0.01$ we plot here the
ratio $d=D_{zx}^2/(D_{zz}D_{xx})$ as a function of position. }\label{Dzx}
\end{figure}

Since the diffusion coefficients, just as the friction coefficients, scale as
the square of a gradient, the largest component is $D_{zz}$. Off-diagonal
elements such as $D_{xz}$ and $D_{zx}$ are, again, smaller by roughly a
factor $kw_0\approx 150$, while the diagonal radial components such as
$D_{xx}$ are in fact largely determined by spontaneous emission, and are of
similar or larger magnitude than the off-diagonal elements. The proper way to
take into account the off-diagonal elements of the diffusion tensor $D$ is to
diagonalize $D$, and consider 3 independent diffusion processes along the
axes of the basis that diagonalizes $D$ with the eigenvalues of $D$ as
diffusion coefficients. Using the fact that $D_{zz}$ is large we can
calculate both eigenvalues and eigenbasis perturbatively. The eigenvalues to
first order are  given by
\begin{eqnarray}
D_{x'x'}&=&D_{xx}-\frac{D_{xz}D_{zx}}{D_{zz}}+\ldots\nonumber\\
 D_{z'z'}&=&D_{zz}+\frac{D_{xz}D_{zx}}{D_{zz}}+\ldots,
\end{eqnarray}
where the $\ldots$ stands for terms of higher order in $1/(kw_0)$, while the
axes change as
\begin{eqnarray}
\hat{z}'&=&\hat{z}+\hat{x}\frac{D_{xz}}{D_{zz}}+\ldots\nonumber\\
 \hat{x}'&=&\hat{x}+\hat{z}\frac{D_{zx}}{D_{zz}}+\ldots.
\end{eqnarray}
The fact that $\hat{z}'$ is slightly tilted towards the $x$ direction implies
that a small part of the large diffusion coefficient $D_{z'z'}$ will
contribute to diffusion in the $x$ direction. This increase, however, is
almost exactly compensated for by the decrease in $D_{x'x'}$. In particular,
the velocity in the $x$ direction undergoes the following Wiener process:
\begin{eqnarray}
{\rm
d}v_x=\sqrt{2D_{x'x'}+2D_{z'z'}\frac{D_{zx}^2}{D_{zz}^2}+\ldots}\,\,\,{\rm
d}W.
\end{eqnarray}
In our case it turns out that $D_{xx}D_{zz}\gg D_{zx}^2$ (see
Fig.~\ref{Dzx}), so that effects due to the off-diagonal elements of the
diffusion tensor can in fact be neglected. The figure also shows that the previous considerations about the relative sizes of the various components of $D$ do not just hold on average, but also locally.
Thus, friction is appreciable only along the cavity axis, while diffusion has
two main contributions: from spontaneous emission in all three directions,
and a large diffusion along the cavity axis from fluctuations in the FORT and
cavity QED forces.
\section{Numerical results}
The following results pertain to a Cs atom, with the ground state given by
$|6S_{1/2}; F=4; m_F=4\rangle$ and the excited state by $|6P_{3/2}; F=5;
m_F=5\rangle$, so that $\lambda_0=852.4$nm and $\Gamma/(2\pi)=5.2$MHz. The
cavity parameters are $\kappa/(2\pi)=4$MHz and $g_0/(2\pi)=30$MHz, and $w_0=20\mu$,
which are
typical for the experiments discussed in \cite{ye}. Furthermore, the values
for $S_0$ examined here are $S_0/(2\pi)=10,50$MHz. Both of these values are
close to those explored in the actual experiment \cite{ye}, and they contrast
the behavior of atoms in shallow ($S_0<g_0$) and deep ($S_0>g_0$) wells.
Typical values for $N_e$ range from $10^{-3}$ to 0.1.
\subsection{Dressed state structure}
\begin{figure}[tbp]
 \leavevmode    \epsfxsize=8cm
\epsfbox{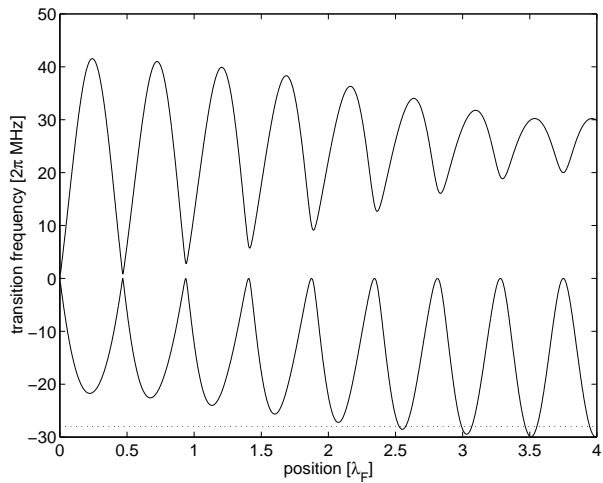} \caption{Transition frequencies $\Delta_{\pm}$ relative to
the bare atomic frequency from the ground state to the lower two excited
dressed states as functions of the position of the atom along the cavity axis
(i.e. $\rho=0$) . Here $S_0/(2\pi)=10$MHz. Also indicated by the dotted line
is the probe detuning used in Fig.~\ref{fric1}, $\Delta_p/(2\pi)=-28$MHz.}
\label{en10}\end{figure}

\begin{figure}[tbp]
 \leavevmode    \epsfxsize=8cm
\epsfbox{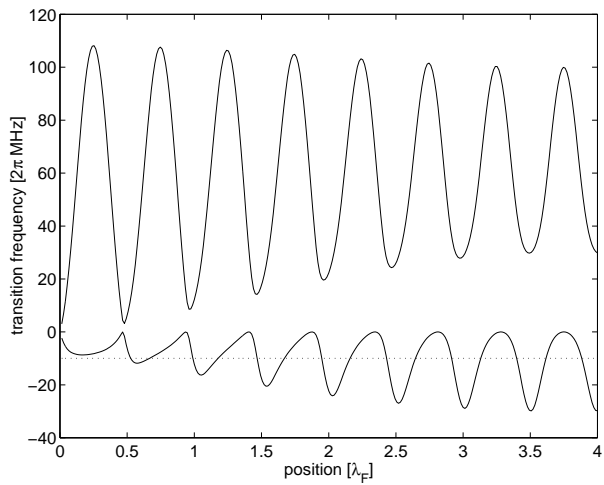} \caption{Same as previous figure but for
$S_0/(2\pi)=50$MHz. Also indicated by the dotted line is the probe detuning
used in Figure~\ref{fric3}, $\Delta_p/(2\pi)=-10$MHz. }\label{en50}
\end{figure}

 We first focus on the atomic motion
along the cavity axis. The simplest way to get a feeling for the results for
$\beta_{zz}$ and $D_{zz}$ as a function of the probe detuning $\Delta_p$ is
to first consider the eigenenergies of the dressed atom-cavity states. When
we neglect dissipation for the moment, and take the limit of no driving
($N_e=0$), we can easily find the energies of the lower dressed states
$|\psi_{\pm}\rangle$ containing at most one excitation: the state containing
no excitation is the ground state with an energy of $E_0=-\hbar
S_F(\vec{r})$, while the energies of the two dressed states in the manifold
of states containing a single excitation are
\begin{equation}
E_{\pm}=\hbar\omega_a\pm \hbar\sqrt{g(\vec{r})^2+S_F(\vec{r})^2},
\end{equation}
 if the atom and cavity are on resonance. The excited dressed states are given by
\begin{equation}
|\psi_-\rangle=\sin\theta |g,1\rangle +\cos\theta |e,0\rangle,
\end{equation}
with
\begin{eqnarray}
\sin\theta&=&\frac{g}{\sqrt{g^2+(\sqrt{g^2+S_F^2}-S_F)^2}},\nonumber\\
\cos\theta&=&\frac{S_F-\sqrt{g^2+S_F^2}}{\sqrt{g^2+(\sqrt{g^2+S_F^2}-S_F)^2}}.
\end{eqnarray}
In Figures~\ref{en10} (10MHz FORT) and \ref{en50} (50MHz FORT) we plot the
transition frequencies (relative to $\omega_a$) from the ground state to
these two excited states as functions of position, i.e.,
\begin{equation}
\Delta_{\pm}=S_F(\vec{r})\pm\sqrt{g(\vec{r})^2+S_F(\vec{r})^2}.
\end{equation}
This expression along with the figures explicitly shows that the main
features of the atom-cavity system are determined by the ratio $S_0/g_0$. It
furthermore shows an important difference with the situation of trapping with
a FORT in free space.  The fact that the excited state shifts {\em up} while
the ground state shifts {\em down} implies that ground and excited states are
trapped in different positions in free space. In the presence of the
quantized cavity field, however, both the lower excited dressed state and the
ground state are now shifted {\em down}. This may improve trapping and
cooling conditions, as detailed below.
\subsection{Cooling mechanisms}
We now take a closer look at cooling mechanisms. In the regime of weak
driving, we will find that the friction coefficient $\beta_{zz}$ is positive
(corresponding to cooling) when the probe field is tuned slightly (by an
amount $~\sim \kappa,\Gamma/2$) below the transition to the relevant dressed
state while for blue detuning the friction coefficient is negative, leading
to exponential heating of the atom's velocity. This can be understood by
analogy with Doppler cooling: by tuning below resonance, the process of
stimulated absorption followed by spontaneous emission leads to a loss of
energy, while the maximum cooling rate is achieved by maximizing the product
of excitation rate and detuning. Now looking back to Figs.~\ref{en10} and
\ref{en50} one sees that the variation of $\Delta_+$ with position is larger
than that of $\Delta_-$, because both the ground state and the lower excited
dressed state shift down, while the upper excited dressed state shifts
upward. Generally speaking, for cooling purposes it is better to tune to the
{\em lower} excited state so as to have smaller spatial variations in cooling
rates. More importantly, the upper excited state energies decrease with
increasing radial distance, whereas the lower excited state energy increases.
Thus, for the upper state the probe detuning changes from red to blue, so
that an atom cooled on axis will in fact be heated if it moves away radially.
For the lower dressed state the probe detuning becomes more red, so that an
atom that is optimally cooled on axis will still be cooled away from the
axis, but at a lower rate.

\begin{figure}[tbp]
 \leavevmode    \epsfxsize=8cm
\epsfbox{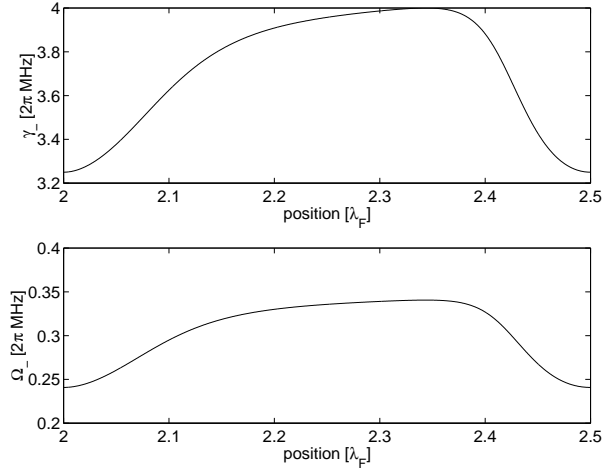} \caption{Decay rate and excitation rate of the lower
dressed state as functions of position along the cavity axis ($\rho=0$). Here
and in Figures \ref{dressed2} and \ref{dressed3}, we chose the following
parameters: $N_e=0.001$, $S_0/(2\pi)=50$MHz,
$\Delta_p/(2\pi)=-10$MHz.}\label{dressed1}
\end{figure}

\begin{figure}[tbp]
 \leavevmode    \epsfxsize=8cm
\epsfbox{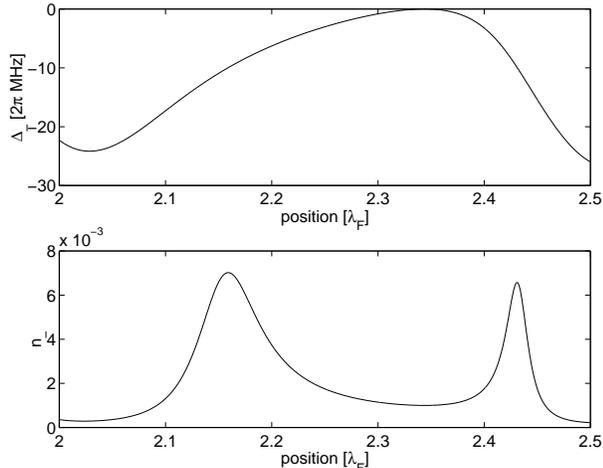} \caption{Transition frequency to and populations in
the lower dressed state as functions of position along the cavity axis. Note
that the equilibrium position of the atom is around
$z=2.25\lambda_F$.}\label{dressed2}
\end{figure}

\begin{figure}[tbp]
 \leavevmode    \epsfxsize=8cm
\epsfbox{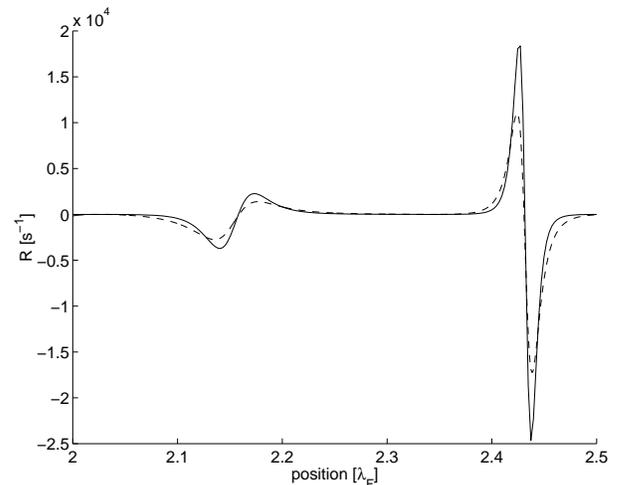} \caption{Cooling rate $\beta_{zz}$ (solid curve) and
the product of gradients of dressed-state population and transition frequency
$R$ (dashed curve)  as functions of position along the cavity axis. The
similarity between the two curves confirms the validity of the Sisyphus
cooling picture. }\label{dressed3}
\end{figure}

The most popular explanation for intra-cavity cooling \cite{horak} exploits
analogies with Sisyphus cooling \cite{sis}, although another explanation for
cavity-based cooling based on asymmetries in coherent scattering was recently
put forward in \cite{chu}. Here we illustrate the Sisyphus picture for
cooling inside optical wells within an optical resonator, using a very simple
dressed-state picture, that makes use of only the lower dressed state and the
ground state, relevant in the low-excitation limit.
 We
choose one particular well, from $z=2.0\lambda_F$ to $z=2.5\lambda_F$, and
one particular set of parameters given in the caption of Fig.~\ref{dressed1}.
In that figure we plot the decay rate $\gamma_-$ of the lower dressed state
and the excitation rate from ground to the dressed state, $\Omega_-$, as
functions of position. In the weak driving limit the decay rate is given by
\begin{eqnarray}
\gamma_-&=&\langle \psi_-|\kappa
a^+a+\Gamma\sigma^+\sigma^-/2|\psi_-\rangle\nonumber\\ &=& \sin^2\theta\kappa
 +\cos^2\theta\Gamma/2,
\end{eqnarray}
and the excitation rate by
\begin{eqnarray} \Omega_-&=&|\langle g,0| {\cal
E}_0(a^++a)|\psi_-\rangle|\nonumber\\ &=& {\cal E}_0|\sin\theta|.
\end{eqnarray}
These two quantities, together with the detuning of the probe field from
(dressed-state) resonance determine the steady-state population in the lower
dressed state, according to
\begin{equation}
n_-=\frac{\Omega_-^2}{(\Delta_--\Delta_p)^2+\gamma_-^2}.
\end{equation}
The population $n_-$ is plotted in Fig.~\ref{dressed2}, along with the
transition frequency $\Delta_-$. These two quantities are sufficient to
understand the Sisyphus cooling mechanism.
Since an atom in the ground state is moving in a conservative potential
around the equilibrium position $z=2.25\lambda_F$, the following Sisyphus
picture should be taken as to apply to the motion of the atom in addition to
that conservative motion (see (\ref{sisf})). Suppose, for example, that the
atom is at position $z=2.2\lambda_F$ and moving towards the right (cf.
Fig.~\ref{dressed2}). The probability to be in the excited state now
decreases (according to the lower part of Fig.~\ref{dressed2}), while the
energy of the excited state relative to the ground state is increasing: in
other words, an atom in the excited state is climbing uphill (again, in
relation to the ground state), but will likely make the down transition to
the ground state, thus leading to cooling at that particular position.
Similarly, at $z=2.4\lambda_F$ an atom moving to the left is going uphill
while having an increased chance of decaying to the ground state, again
leading to cooling. This picture in fact shows that the cooling rate is
expected to be proportional to the gradient of $n_-$ and the gradient of
$\Delta_-$. More precisely, the force on the atom at position $z$ is
approximately given by
\begin{eqnarray}\label{sisf}
F_z&\approx&\hbar\frac{d S_F}{dz}-\hbar
n_-(z-v/\gamma_-)\frac{d\Delta_-}{dz}\nonumber\\ &\approx& \hbar\frac{d
S_F}{dz}-\hbar n_-(z)\frac{d\Delta_-}{dz} + \frac{\hbar
v}{\gamma_-}\frac{dn_-}{dz}\frac{ d\Delta_-}{dz},
\end{eqnarray}
where the argument of $n_-$ indicates the lag between the atom reaching a
position $z$ and reaching its steady state, with the lag time scale
determined by the inverse decay rate from the dressed state. From the second
line we see that the friction coefficient $\beta_{zz}$ is approximated by
\begin{equation}
R\equiv-\frac{\hbar}{m \gamma_-}\frac{dn_-}{dz}\frac{ d\Delta_-}{dz}.
\end{equation}
Indeed, Fig.~\ref{dressed3} shows the similar behavior of $\beta_{zz}$ and
$R$ as functions of position.
\subsection{Friction, diffusion and equilibrium rms velocities}\label{cool}
\begin{figure}[tbp]
 \leavevmode    \epsfxsize=8cm
\epsfbox{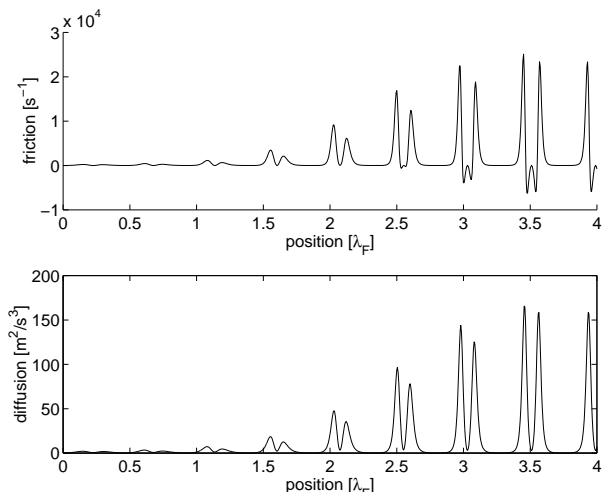} \caption{Friction and velocity diffusion
coefficients $\beta_{zz}$ and $D_{zz}$ as functions of the atomic position
(in units of $\lambda_F$) along the cavity axis. Here $N_e=0.001$,
$S_0/(2\pi)= 10$MHz, and $\Delta_p=-28\times 2\pi$MHz. Cf.\
Fig.~\ref{en10}.}\label{fric1}
\end{figure}

\begin{figure}[tbp]
 \leavevmode    \epsfxsize=8cm
\epsfbox{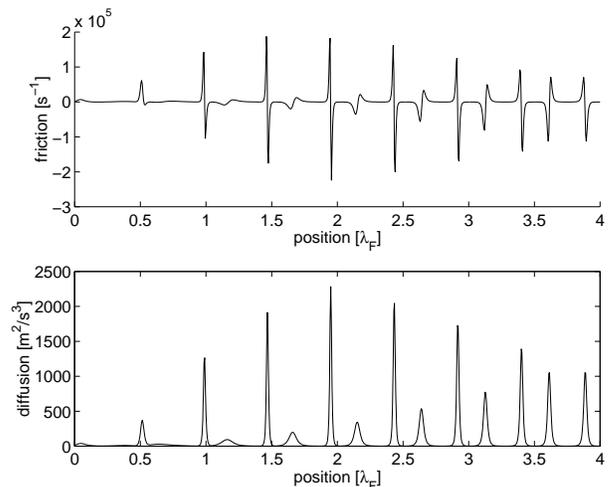} \caption{Same as previous Figure but for
$N_e=0.01$, $S_0/(2\pi)= 50$MHz, and $\Delta_p=-10\times
2\pi$MHz.}\label{fric3}
\end{figure}

 In Figures~\ref{fric1}--\ref{fric3} we give examples of friction
and diffusion coefficients for both the 10 and 50 MHz FORTs, as functions of
the atomic position.  They illustrate the point that in the low-excitation
limit red (blue) detuning leads to cooling (heating) (cf. Figs.~\ref{en10}
and \ref{en50}). They moreover clearly show how all wells are quantitatively
different, with cooling rates and diffusion strengths differing by orders of
magnitude over the various wells, and with $\beta_{zz}$ being negative in
some wells, and always positive in others. This of course also implies that
the temperatures reached by atoms in thermal equilibrium vary with position.

\begin{figure}[tbp]
 \leavevmode    \epsfxsize=8cm
\epsfbox{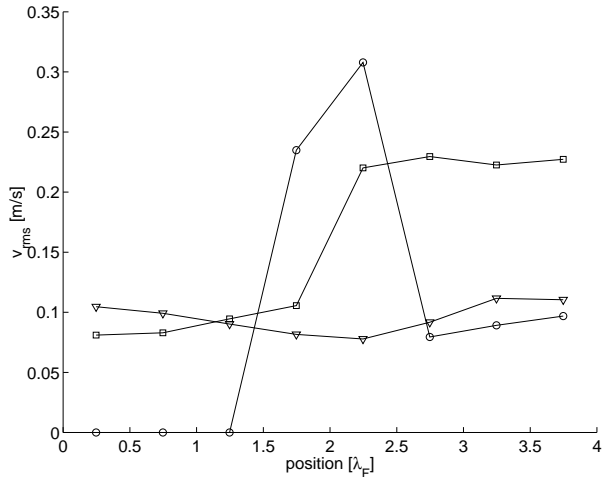} \caption{The values of $v_{{\rm rms}}^z$ in the eight
equilibrium points as defined in (\protect{\ref{vrmsa}}) by averaging over
the entire well. In all cases $S_0=2\pi\times 10$MHz. Triangles correspond to
a detuning $\Delta_p/(2\pi)=-28$MHz, squares to $\Delta_p/(2\pi)=-23$MHz, and
circles to $\Delta_p/(2\pi)=15$MHz . Note the points on the latter curve on the axis 
indicate that the friction coefficient is negative, so that
there is in fact no cooling and $v_{{\rm rms}}^z$ is not
defined. They do not indicate cooling to $v_{{\rm rms}}=0$.}\label{vrms1}
\end{figure}

\begin{figure}[tbp]
 \leavevmode    \epsfxsize=8cm
\epsfbox{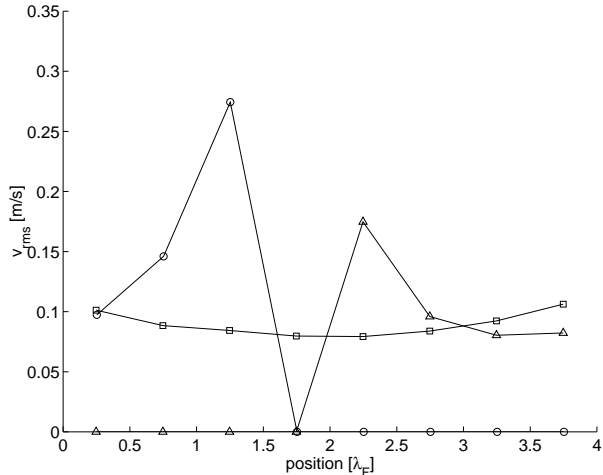} \caption{As previous Figure, but for $S_0=2\pi\times
50$MHz and $N_e=0.01$, and where the average is taken over a region of size
$\lambda_F/10$ around the equilibrium point. The probe detunings were
$\Delta_p/(2\pi)=-10,-5,100$MHz for the squares, triangles, and circles,
respectively. 
Note the points on the axis 
indicate that the friction coefficient is negative, so that
there is in fact no cooling and $v_{{\rm rms}}^z$ is not
defined. They do not indicate cooling to $v_{{\rm rms}}=0$.
 }\label{vrms2}
\end{figure}

For the case of the shallow FORT we consider weak driving ($N_e=0.001$),
whereas for the deeper FORT the driving field is taken to be stronger by an
order of magnitude. The stronger driving field increases cooling rates while
the fact that deeper wells trap the atoms better means that correspondingly
larger diffusion rates still can be tolerated.
 The stable equilibrium points $z^e_n$ are located around
the maxima of $S_F$, i.e. around $z_n=(n-1/2)\lambda_F/2$ for integer $n$,
because it is the FORT that gives the main contribution to the total force
(even for the smallest value of $S_0=2\pi\times 10$MHz considered here). The
cavity QED field gives only a small correction to the force and hence to the
equilibrium position. In each equilibrium point, we can define a measure for
the expected rms velocity of the atom along the $z$ axis  in thermal
equilibrium by considering averages over local wells
\begin{equation}\label{vrmsa}
v_{{\rm rms}}^z= \sqrt {\frac{\bar{D}_{zz}}{\bar{\beta}_{zz}}} \,\,{\rm if}\,
\bar{\beta}_{zz}>0,
\end{equation}
in terms of the friction and diffusion coefficients. This averaging procedure
gives a sensible measure for the rms velocity only if the atom indeed samples
the whole well. This condition is fulfilled for the relatively shallow wells
originating from $S_0=2\pi\times 10$MHz, and Fig.~\ref{vrms1} uses this
averaging procedure. For the 50MHz FORT, however, we averaged over only part
of the well, namely a region of size $\lambda_F/10$ symmetrically around the
equilibrium point. This choice is rather arbitrary, and thus Fig.~\ref{vrms2}
just gives an indication of what rms velocities to expect for atoms trapped
in the corresponding wells, although the simulations in fact do confirm these
values.
 We see here that depending on the probe detuning, the
atom will be cooled to low temperatures either in all wells, or only in wells
where $g$ is large in the equilibrium point, or only in wells where $g$ is
small. This shows the flexibility that a FORT beam adds: one can predetermine
to a certain degree in which well the atom will be trapped (and cooled) for
longer times and in which it will not be.

Under the current conditions $\kappa>\Gamma/2$ the lowest temperatures
achievable are determined by the Doppler velocity $v_D$. More precisely, the
lower limit on rms velocities along the cavity axis is expected to be
\begin{equation}
v_D^z=\sqrt{0.7\frac{\hbar\Gamma}{2m}},
\end{equation}
where the factor $0.7=(1+2/5)/2$ comes from the fact that in our case the
diffusion due to spontaneous emission in the $z$ direction is 2/5th of the
full 3-D value.
 We tested that for smaller
$\kappa$ the rms velocities indeed do become even smaller, now determined by
$\sqrt{\hbar\kappa/m}$, thus confirming predictions of \cite{horak}.

Finally, we note that the quasiclassical approximation used throughout this
paper is justified as neither the recoil limit is reached nor the
resolved-sideband limit, i.e.
\begin{eqnarray}
\hbar\Gamma/2&\gg& (\hbar k)^2/m,\nonumber\\
 \hbar\Gamma/2&\gg& h \nu_{{\rm osc}},
\end{eqnarray}
with $\nu_{{\rm osc}}$ the oscillation frequency of the atom in a well (see
below), although in some cases the latter condition is only marginally
fulfilled, namely when $\nu_{{\rm osc}}=600$kHz, which is only a factor 4
smaller than $\Gamma/(4\pi)$.
\subsection{Saturation behavior} 
\begin{figure}[tbp]
 \leavevmode    \epsfxsize=8cm
\epsfbox{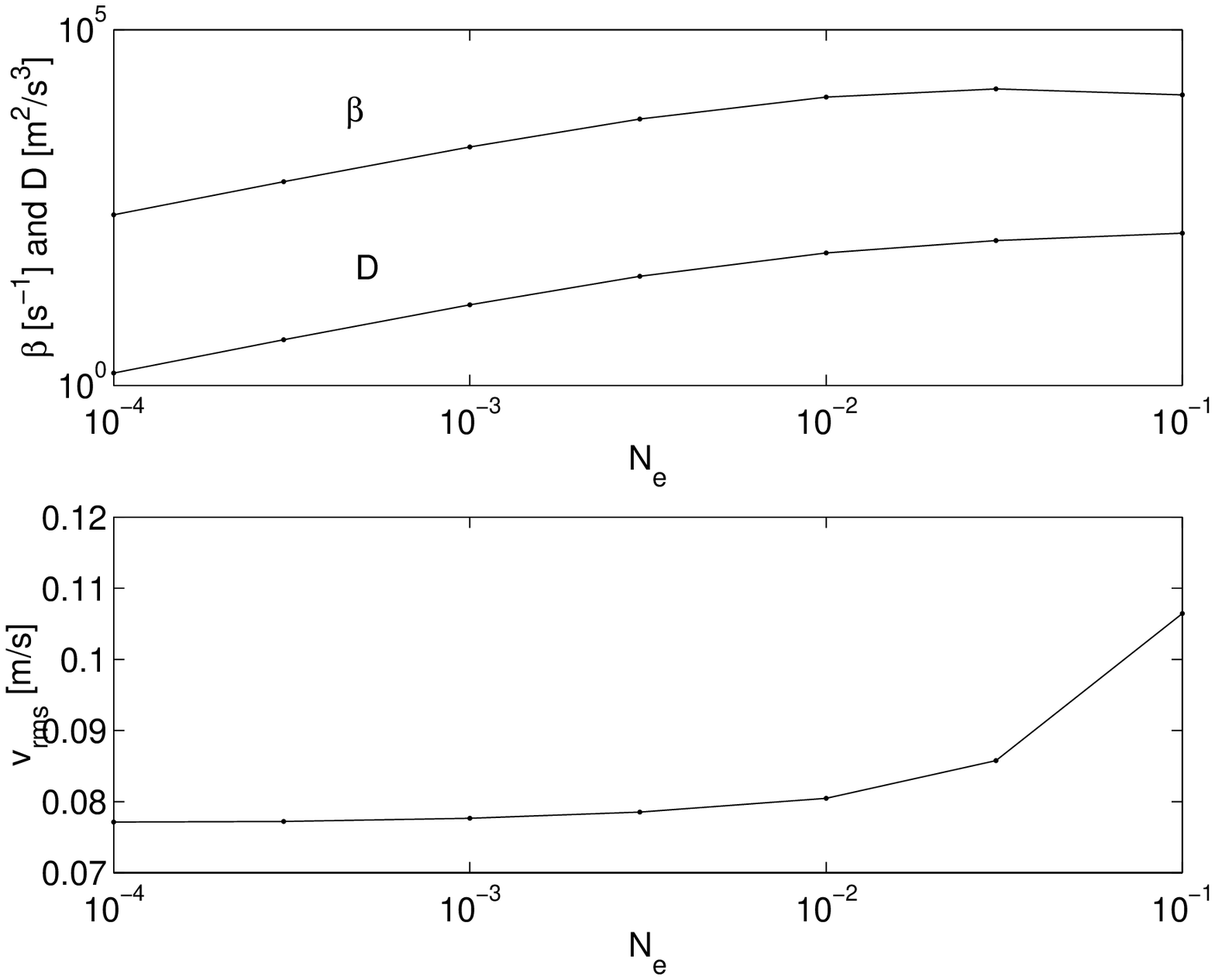} \caption{The average values of $D_{zz}$  and $\beta_{zz}$
as functions of the driving field strength $N_e$ in the well extending from
$z=2\lambda_F$ to $z=2.5\lambda_F$, for the 10 MHz FORT, where
$\Delta_p=-28\times 2\pi$MHz. In the lower part the corresponding values for
the rms velocity $v_{{\rm rms}}^z$ are plotted as a function of
$N_e$.}\label{scale1}\end{figure}

We now briefly turn to the
question of the nonlinear behavior of the atom-cavity system with increasing
excitation. In the absence of saturation effects, both friction and diffusion
coefficients would increase linearly with $N_e$. For the same parameters as
Fig.~\ref{fric1}, Figure~\ref{scale1} shows nonlinearities setting in around
$N_e=0.01$. The friction coefficient even starts to {\em decrease} around
$N_e=0.1$ as a result of the local values of $\beta_{zz}$ becoming negative
where they were positive in the weak driving limit. The concomitant effect on
the $v_{{\rm rms}}^z$ is shown as well.
\subsection{Simulations}\label{sims}
We also performed Monte-Carlo simulations of the 3-D motion in given wells by
solving the Langevin equations (\ref{lang}) for position and velocity (see
also \cite{doherty}). The experimental procedure switches the FORT field on
only when an atom has been detected and when it consequently has partly
fallen through the cavity already \cite{ye}. We accordingly fix initial
conditions as follows: We start the atom on the cavity axis, and we fix the
downward velocity to be $v_x=10$ cm/s. Furthermore, we chose $v_z=0$ cm/s,
and the initial position along the $z$ axis to be $\lambda_F/8$ away from the
equilibrium point.  The initial position and velocity were fixed so that all
variations in trapping times and rms velocities are solely due to the random
fluctuations of the forces acting on the atom, rather than from random
initial conditions. Experimentally these two are mixed of course. 

Since atoms with these initial conditions do not possess angular momentum around the $z$ axis, this in some sense represents a favorable case (although the atoms are not put in the bottom of the well). However, in the course of their evolution the atoms do acquire angular momentum  so that this is in fact not a severe restriction. For more detail see below (Figure~\ref{init}).

\begin{figure}[tbp]
 \leavevmode    \epsfxsize=8cm
\epsfbox{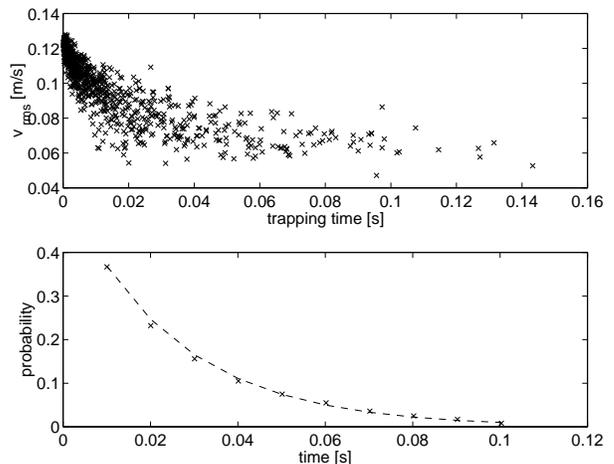} \caption{In the upper part of the plot each data point
gives the trapping time and the average $v_{{\rm rms}}$ resulting from a
single trajectory. Identical initial conditions were chosen for each
trajectory: each atom started at $z=2.125\lambda_F$ with $v_z=0$ and
$v_x=10$cm/s (downward). Other parameters were: $N_e=0.001$,
$\Delta_p=-28\times 2\pi$MHz, $S_0=10\times 2\pi$MHz. The lower part gives a
histogram of the probability $P(T)$ for an atom to be trapped longer than a
time $T$. A fit of the tail of this distribution to an exponential $\propto
\exp(-T/\tau)$  gives $\tau=(25\pm 2)$ ms } \label{sim1h}\end{figure}

In Figure~\ref{sim1h} we plot the results of simulations of  1000
trajectories for an atom in the shallow well of 10 MHz. We plot the average
rms velocity along the cavity axis as a function of trapping time for each
trajectory. Here we defined the ``trapping time'' as the time spent by the
atom in one particular given well of size $\lambda_F/2$. The actual trapping
time inside the cavity may be longer, obviously, as the atom may subsequently
get trapped in different wells.  For very short trapping times, $v_{rms}$ is
determined by the initial condition, but for longer times lower temperatures
corresponding to those calculated in Fig.\ref{vrms1} are reached. Note
however that the simulations were done in 3-D, and as such do not necessarily
give the same temperatures as predicted for on-axis (1-D) motion in
Figures~\ref{vrms1} and \ref{vrms2}. Nevertheless, the effect of the atoms'
radial motion is apparently not strong, and in fact atoms leave the well
while still being trapped radially. This is partly due to the fact that all
(especially heating) rates  in the radial direction are smaller by a factor
$kw_0\approx 150$ than those in the axial direction.
 About
half of the atoms is basically not trapped at all. The remaining atoms have a
probability $P(T)$ to be trapped longer than a time $T$, with $P(T)$ decaying
exponentially with $T$. The average trapping time for these parameters is
found to be $\tau\approx$ 25ms, as shown in Fig.~\ref{sim1h}.

\begin{figure}[tbp]
 \leavevmode    \epsfxsize=8cm
\epsfbox{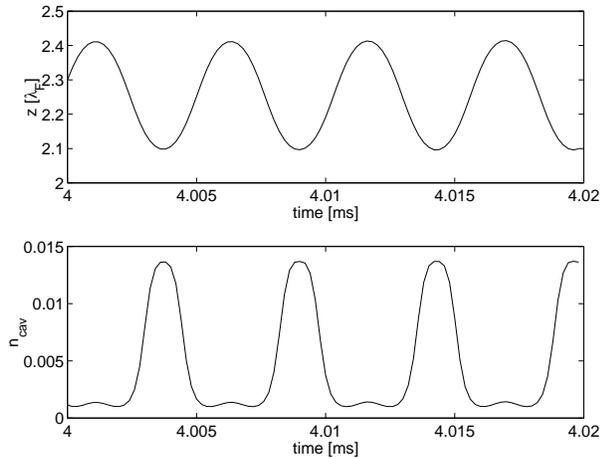} \caption{Snapshot of a single trajectory, with parameters
as in Figure \ref{sim1h}. The upper plot gives the $z$ coordinate of the atom
as a function of time, the lower plot gives the transmission (in fact the
number of photons $\langle a^+a\rangle$ inside the cavity) in that same time
interval. Note the time scales here differ by two orders of magnitude from
those of Figure~\ref{trajrho}.} \label{trajz}\end{figure}

\begin{figure}[tbp]
 \leavevmode    \epsfxsize=8cm
\epsfbox{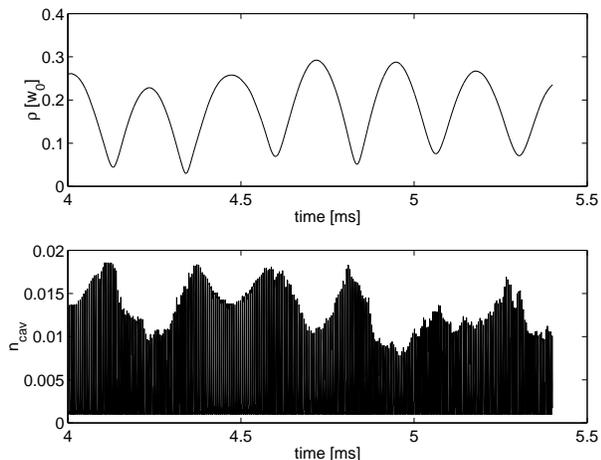} \caption{For the same trajectory as  the previous
figure, the upper plot gives the radial distance to the cavity axis, $\rho$
in units of $w_0$ as a function of time, the lower plot gives again the
number of photons inside the cavity during that same time interval. The atom
has a nonzero angular momentum along $z$ and does not cross the $z$ axis.}
\label{trajrho}\end{figure}

 In Figures~\ref{trajz}
and \ref{trajrho} we plot for the same 10MHz FORT an example of a single
trajectory, after the atom has spent 4ms in the trap. The oscillation
frequencies along the $z$ and the radial directions differ by two orders of
magnitude (since $kw_0\approx 150$): in the $z$ direction the oscillation
rate is $\sim$ 200kHz, in the radial direction $\sim$ 2.2kHz. The photon
transmission follows both these oscillations so that in principle the atomic
motion in both axial and radial direction is detectable. Experimentally,
though, the oscillations along the cavity axis may be too fast to be
accessible. In particular, the average rate at which photons leaking out
through one end of the cavity are detected is at most (the efficiency is less
than 100\%) equal to the cavity decay rate multiplied by the average number
of photons inside the cavity. For the parameters of Fig.~\ref{trajz} this
amounts to a rate $\sim 0.01\times\kappa\approx 2.5\times 10^5$/sec, which
corresponds to just about one photon per oscillation period.

The figures show that when the atom is in a position where it is not coupled
to the cavity ($g=0$), the number of photons in the cavity drops to
$N_e=0.001$. Similarly, when the atom moves away radially, the transmission
drops.
 To
make a direct comparison with the trapping times achieved in the experiment
\cite{ye}, we now turn to the case of a 50 MHz FORT. We plot rms velocities
vs trapping times for 300 trajectories for an atom trapped in the well
ranging from $z=2\lambda_F$ to $z=2.5\lambda_F$.
\begin{figure}[tbp]
 \leavevmode    \epsfxsize=8cm
\epsfbox{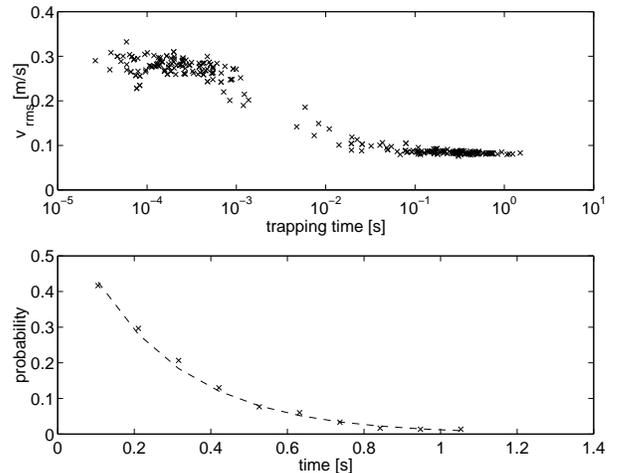} \caption{As Fig.~\ref{sim1h} but for $N_e=0.01$,
$\Delta_p=-10\times 2\pi$MHz, $S_0=50\times 2\pi$MHz. The mean trapping time
is $\tau=250\pm20$ms. } \label{sim2h}\end{figure} For the parameters of
Fig.~\ref{sim2h} the atom is either trapped for long times ($>10$ms) or only
for a short time ($<1$ms), both with about $50\%$ probability. In the latter
case the rms velocity is determined just by the (arbitrarily chosen) initial
condition and is around 30cm/s, but for longer trapping times the effects of
cooling are visible. Thermal equilibrium is reached with $v_{rms}\sim 8$cm/s,
thus confirming the results of Fig.~\ref{vrms2}. The distribution of trapping
times again follows an exponential law, and the average trapping time, as
determined from the tail of the distribution, is $\tau\approx 250$ms, which
is ten times longer than for the (fluctuating) 50MHz FORT used in \cite{ye}.
This shows the great potential of holding single atoms in the cavity for
extended periods of time if the intensity fluctuations of the FORT beam can
be minimized. Experimental efforts along this path are currently underway.

 Also for this case we
plot snapshots for a single trajectory, taken after the atom has spent 25ms
in the trap. Compared to the 10MHz FORT, the oscillations of the atom  along
the cavity axis and in the radial direction become faster by about a factor
of 3. The axial oscillation frequency is about 600kHz, while along the radial
direction the oscillations occur at a rate 6.2kHz, i.e., again slower by two
orders of magnitude. In this case, the photon transmission still follows
directly the axial oscillations but no longer follows the radial excursions
of the atom, as now the fluctuations in the magnitude of $g$ at the atom's
position along the cavity axis are in fact larger than those due to the
radial excursions of the atom. This is partly due to the fact that in the
simulations here the driving field is stronger than for the 10MHz example
above so that fluctuations in the atomic motion occur at a shorter time
scale, and partly simply because the radial excursions are small.
Fig.~\ref{trajr2}(c) shows that it is  primarily the axial fluctuations that
determine the variations in the numbers of photons inside the cavity.

Generally speaking, the axial excursions determine (local) minimum and maximum transmission
levels (as in Fig. 15). When these minima and/or maxima depend on the radial 
position, then the radial motion could in principle be visible in the cavity transmission level. 
This depends in turn on whether the axial fluctuations on the time scale of 
the transverse motion are sufficiently small so as not to hide the radial dependence.
There seems to be no simple general rule how this interplay between radial and axial motions depends 
on detunings, driving strength, and the particular well.
\begin{figure}[tbp]
 \leavevmode    \epsfxsize=8cm
\epsfbox{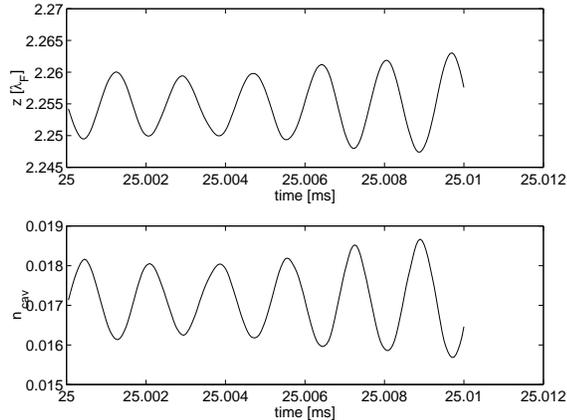} \caption{Snapshot of a single trajectory, with
parameters as in Figure \ref{sim2h}. The upper plot gives the $z$ coordinate
of the atom as a function of time (with the atom released with standard
initial conditions at $t=0$). The lower plot gives the transmission (in fact
the number of photons inside the cavity) in that same time interval. Note the
time scales here differ by two orders of magnitude from those of
Figure~\ref{trajr2}.} \label{trajz2}\end{figure}
\begin{figure}[tbp]
 \leavevmode    \epsfxsize=8cm
\epsfbox{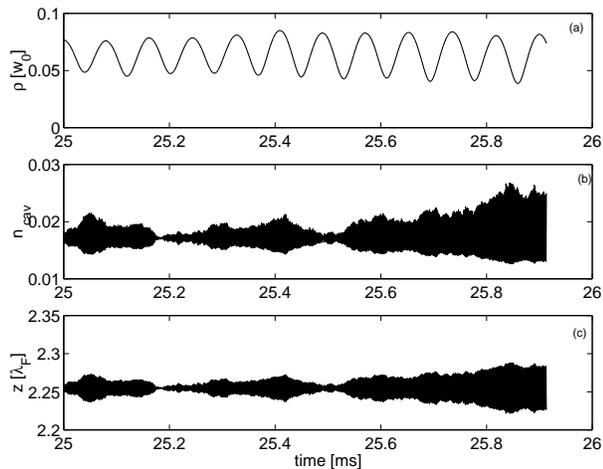} \caption{For the same trajectory as  the previous
figure, (a) the radial distance to the cavity axis, $\rho$ in units of $w_0$
as a function of time, (b) the number of photons inside the cavity during
that same time interval, and (c) the position along the cavity axis in units
of $\lambda_F$.} \label{trajr2}\end{figure} In contrast, in a different well,
the one ranging from $z=0.0$ to $z=0.5\lambda_F$, the photon number in the
cavity does follow the radial motion, as the radial excursions become larger.
Perhaps more importantly, the average transmission level is higher by more
than a factor 2 compared to the previous case, as a result of $g$ being
larger in this well (cf.~Fig.~\ref{beatgF}). This shows how, in principle,
different wells may be experimentally distinguished via the
transmission of the probe field through the cavity.
\begin{figure}[tbp]
 \leavevmode    \epsfxsize=8cm
\epsfbox{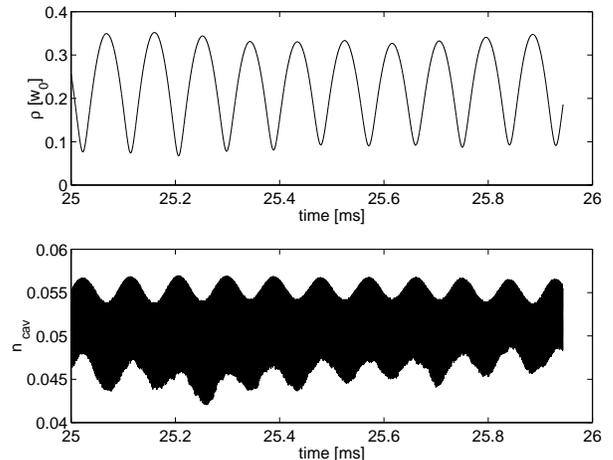} \caption{For an atom in the well ranging from $z=0$ to
$z=\lambda_F/2$, for $N_e=0.01$, $\Delta_p=-10\times 2\pi$MHz, and
$S_0=50\times 2\pi$MHz, the upper plot gives the radial distance to the
cavity axis, $\rho$ in units of $w_0$ as a function of time, the lower plot
gives the number of photons inside the cavity during that same time
interval.} \label{trajr3}\end{figure} We also simulated the motion of an
atom trapped under more adverse conditions, namely for an atom in the well
$[z=\lambda_F\rightarrow 1.5\lambda_F]$ at a probe detuning
$\Delta_p/(2\pi)=-5 $MHz. According to Fig.~\ref{vrms2}, the atom is not
cooled on axis under these conditions (i.e. the average friction coefficient
around the equilibrium point on the $z$ axis is negative). This is confirmed
by Fig.~\ref{simbad}: the mean trapping time for an atom starting at
$z=1.125\lambda_F$ is now very short, about 1.6ms, while the average rms
velocity is  $v_{{\rm rms}}^z\approx 28$ cm/s, as determined essentially by
the initial condition.
\begin{figure}[tbp]
 \leavevmode    \epsfxsize=8cm
\epsfbox{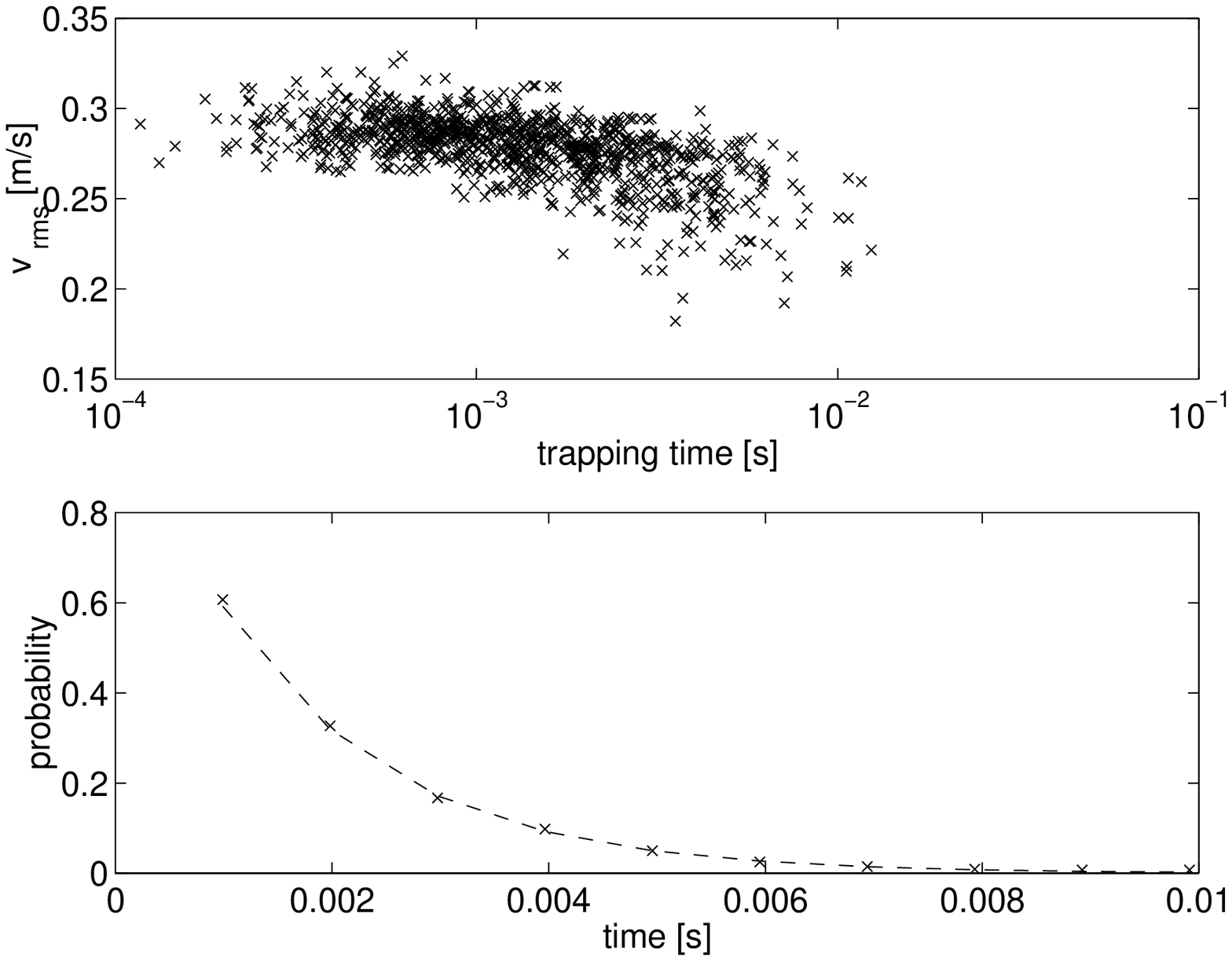} \caption{As Fig.~\ref{sim1h} but for $N_e=0.01$,
$\Delta_p=-5\times 2\pi$MHz, $S_0=50\times 2\pi$MHz. The initial position of
the atom is $z=1.125\lambda_F$. The mean trapping time is $\tau=1.6\pm
0.1$ms. } \label{simbad}\end{figure}

Finally, we consider the influence of different initial conditions on trapping and cooling. All the results so far were obtained by considering atoms that initially are moving on axis. Thus, they have no angular momentum along the $z$ axis, nor any radial potential energy. 
Figure~\ref{init} shows a plot of rms velocities vs. trapping times for atoms trapped under the same conditions as for Fig.~\ref{sim2h}) (i.e., in the well from $z=2\lambda$ to $z=2.5\lambda$, for $\Delta_p=-10\times 2\pi$MHz, $N_e=0.01$, and $S_0=50\times 2\pi$MHz),
but with different (nonzero) values for the initial angular momentum.   
\begin{figure}[tbp]
 \leavevmode    \epsfxsize=8cm
\epsfbox{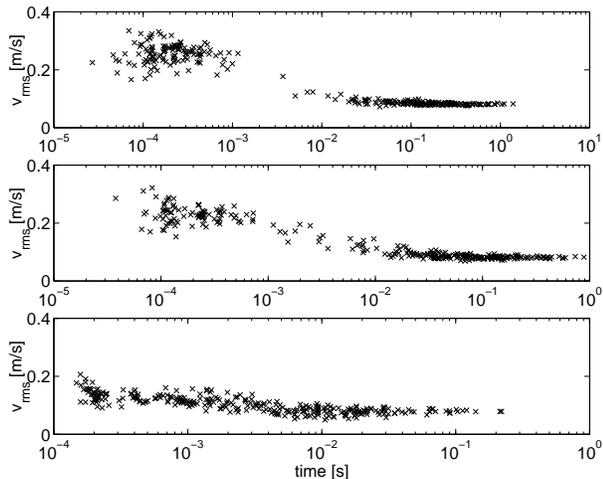} \caption{Rms velocities vs. trapping times
for atoms trapped under the same conditions as for Fig.~\ref{sim2h} but with different initial radial conditions for $y$. In particular, for plot (a) the initial conditions on $y$ is $y=0.2w_0$.
for (b) $y=0.5w_0$ and for (c) $y=w_0$. Since $v_x=-10$cm/s, the atoms have different angualr momenta along $z$ in these cases, and different initial potential energies.} \label{init}\end{figure}
Obviously, the more initial potential energy the atom has, the less likely it is to be trapped. In fact, the angular momentum does not play any role here, as confirmed by similar calculations with initial conditions chosen such that the atoms have no initial angular momentum but have the same potential energy. The results are the same in that case.
For atoms starting at $y=0.2w_0$ the trapping times and rms velocities are basically not affected, and the trapping time is still around 250ms. But for atoms starting at $y=0.5w_0$ the effect of their increased potential energy leads to clearly shorter trapping times (by roughly a factor of 2), and for atoms starting at $y=w_0$ this effect is even more pronounced with a decrease in trapping time of about a factor of 10.

\subsection{A different trapping structure}\label{caseB}
We now consider a different case where the atomic excited state is assumed to
be shifted {\em down} by the FORT field, just as the ground state is (see,
for instance \cite{icols}). This can be achieved by using a FORT that is
(red) detuned in such a way that the excited atomic state is relatively
closer to resonance with a higher-lying excited state than with the ground
state. This situation at first sight looks even more appealing for trapping
purposes, as now both excited and ground state will be trapped in the same
positions. Moreover, fluctuations in the force due to the FORT are dimished.

We consider only the 50MHz FORT here, and compare this case to the previous
50MHz FORT case, and in particular we refer the reader back to
Figs.~\ref{en50}, \ref{fric3}, and \ref{sim2h}. For ease of comparison we
keep $\lambda_F$ the same, and assume for simplicity that the excited state
is shifted down by an amount $S_F$, so that the shifts of the ground and
excited state are in fact identical.

The fact that ground and excited state see the same potential, implies that
the transition frequencies to the dressed states are simply periodic in space
with period $\lambda_0$, as shown in Fig.~{\ref{en50B}, rather than aperiodic
as in Fig.~\ref{en50}.
\begin{figure}[tbp]
 \leavevmode    \epsfxsize=8cm
\epsfbox{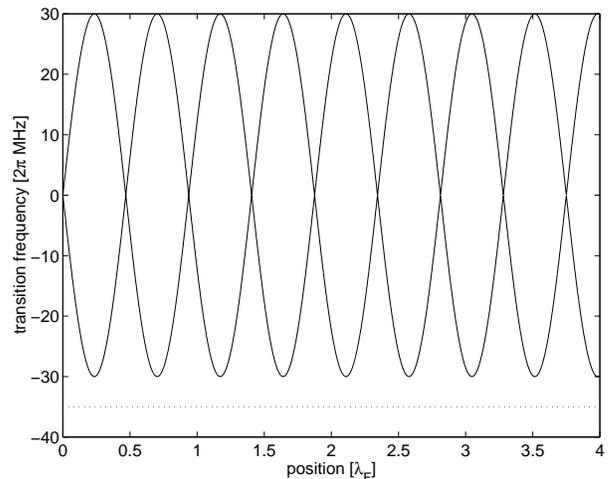} \caption{Transition frequencies $\Delta_{\pm}$ for the
case where the atomic excited state is assumed to be shifted down by the FORT
field by the same amount $S_F$ as is the ground state, for the 50MHz FORT. A
detuning of $-35$MHz is indicated by the dotted line.}\label{en50B}
\end{figure}
Similarly, the fluctuations in the force due to the FORT now vanish, as both
ground and excited state undergo the same shift, so that the diffusion
coefficient is periodic with period $\lambda_0$. Also the friction force
arises only from the cavity QED part and is periodic. Yet, the different
wells are not equivalent. The forces are, just as before, driven by both
cavity QED field and the FORT, and the value of $g$ at the antinode of the
FORT still varies over the different wells. This is illustrated in
Fig.~\ref{fricB} where the rms velocities in the 8 different wells are shown,
along with friction and diffusion coefficients.
 Since in this example the probe
field is detuned below the lower dressed state, one has cooling everywhere in
space.
\begin{figure}[tbp]
 \leavevmode    \epsfxsize=8cm
\epsfbox{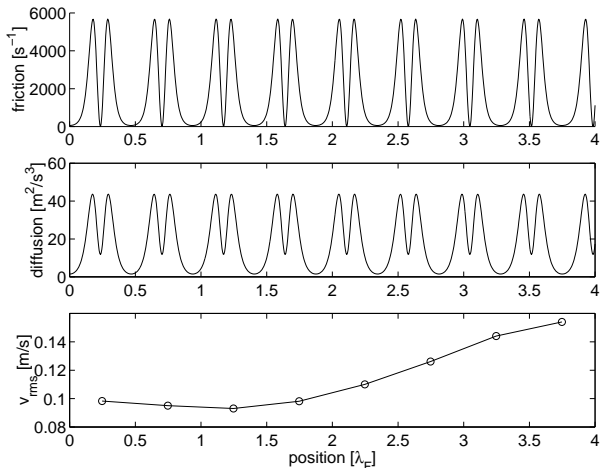} \caption{Friction and diffusion coefficients, and the
resulting rms velocity as functions of position along the cavity axis for the
trapping structure of Section \ref{caseB}. Here $\Delta_p/(2\pi)=-35$MHz, and
$N_e=0.01$. } \label{fricB}\end{figure} The simulations show that the mean
trapping time is smaller, although the rms velocities are just as small as
before. The reason is the less favorable cooling condition away from the
cavity axis. In particular, for the parameters used here the expected rms
velocity $v_{{\rm rms}}^z$ steadily increases to 90cm/s at a radial distance
$\rho=2w_0$, while for the simulations of Figs.~\ref{sim2h}, $v_{{\rm
rms}}^z$ is increasing only slowly to 12cm/s. This large difference can be
understood by noting the difference in dressed state structures between the
two cases. For the case of Fig.~\ref{en50}, the transition frequency to the
lower dressed state around the equilibrium position $z\approx 2.25\lambda_F$
does not change much with increasing radial distance, so that the probe field
in that trapping region is always detuned below resonance by an amount that
stays more or less constant. For the dressed state structure of
Fig.~\ref{en50B}, however, the probe detuning increases from $\geq 5$MHz to
$\geq 35$MHz below resonance, thus leading to much worse cooling conditions.
In other words, the presence of opposite level shifts due to the FORT makes
the spatial variation of the transition frequency to the lower dressed state
{\em smaller}: compare $\Delta_-=S_F-\sqrt{S_F^2+g^2}$ to $\Delta_-=-g$,
especially when $g\ll S_F$.

 The alternative trapping potential is, therefore, not necessarily more
favorable for trapping purposes. On the other hand, {\em all} atoms are
captured now and are trapped for at least 10ms. This can be understood from
the simple fact that here the friction coefficient is positive in the {\em
entire} well.
\begin{figure}[tbp]
 \leavevmode    \epsfxsize=8cm
\epsfbox{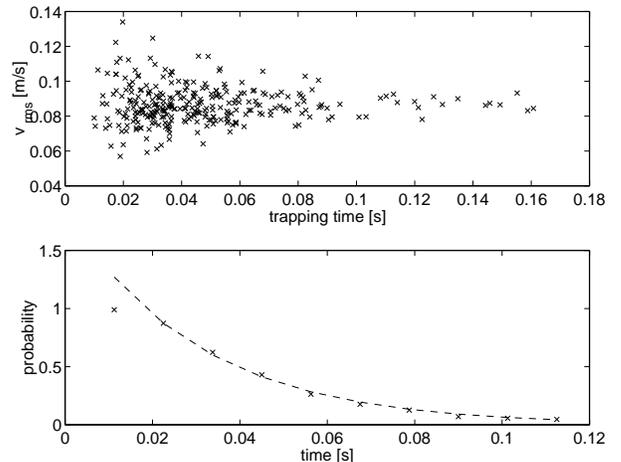} \caption{As Figure~\ref{sim1h}, but for the different
trapping structure of Section \ref{caseB}. The initial position was
$z=0.125\lambda_F$, and further parameters were $N_e=0.01$,
$\Delta_p=-35\times 2\pi$MHz, $S_0=50\times 2\pi$MHz. The mean trapping time
is $\tau=28\pm2$ms. } \label{simBh}\end{figure}
\section{Summary}
We analyzed cooling limits and trapping mechanisms for atoms trapped in
optical traps inside optical cavities. The main distinguishing feature from
previous discussions on cooling of atoms inside cavities is the presence  of
the external trapping potential with a different spatial periodicity as
compared to the cavity QED field. This not only provides better cooling and trapping conditions
but the different spatial period makes the various
potential wells qualitatively different. Atoms can be trapped in regions of
space where the coupling to the cavity QED field is maximum, minimum or
somewhere in between. Depending on the laser detuning, cooling may take place
only in wells where the atom is minimally coupled to the cavity QED field, or
where it is maximally coupled. This allows one in principle to distinguish
to a certain degree the different atomic positions along the cavity axis, namely, by comparing 
\begin{itemize}
\item
the average transmission level 
\item
the fluctuations of the cavity transmission 
\item
the total trapping time
\end{itemize}
which reflect, respectively, the average atom-cavity coupling,
the temperature of the atom and under certain conditions the radial motion, 
and the overall cooling and trapping conditions.  
This is an important additional tool useful for eventual
control of coherent evolution of the atomic center-of-mass degrees of
freedom, as relevant to performing quantum logic operations.
\section*{Acknowledgements}
We thank Andrew Doherty, Klaus M\o lmer and David Vernooy for helpful discussions and comments.
This work was funded by DARPA through  the National Science Foundation and the
QUIC (Quantum Information and Computing) program administered by the US Army
Research Office and the Office of Naval Research.

\end{document}